\documentclass[aps,prl,nofootinbib,superscriptaddress,twocolumn,amsmath,amssymb]{revtex4}
\usepackage{amsmath,amsfonts,amssymb,graphicx}
\usepackage{epsfig}
\usepackage[usenames]{color}
\usepackage{epstopdf}
\newcommand{\be}{\begin{equation}}
\newcommand{\ee}{\end{equation}}
\newcommand{\bey}{\begin{eqnarray}}
\newcommand{\eey}{\end{eqnarray}}
\newcommand{\bw}{\begin{widetext}}
	\newcommand{\ew}{\end{widetext}}

\newcommand{\ba}{\begin{array}}
	\newcommand{\ea}{\end{array}}
\newcommand{\bi}{\begin{itemize}}
	\newcommand{\ei}{\end{itemize}}
\newcommand{\bem}{\begin{enumerate}}
	\newcommand{\eem}{\end{enumerate}}

\newcommand{\hefei}{Department of Modern Physics, University of Science and Technology of China, Hefei 230026, China}
\newcommand{\como}{Center for Nonlinear and Complex Systems, Dipartimento di Scienza e Alta Tecnologia,
	Universit\`a degli Studi dell'Insubria, via Valleggio 11, 22100 Como, Italy}
\newcommand{\infn}{Istituto Nazionale di Fisica Nucleare, Sezione di Milano, via Celoria 16, 20133 Milano, Italy}
\newcommand{\brazil}{International Institute of Physics, Federal University of Rio Grande do Norte,
	Campus Universit\'ario - Lagoa Nova, CP. 1613, Natal, Rio Grande Do Norte 59078-970, Brazil}
\newcommand{\NEST}{NEST, Istituto Nanoscienze-CNR, I-56126 Pisa, Italy}

\begin{document}

\title{Quantum Chaos and the Correspondence Principle}

\author{Jiaozi Wang}
\email{wangjz@mail.ustc.edu.cn}
\affiliation{\hefei}
\author{Giuliano Benenti}
\email{giuliano.benenti@uninsubria.it}
\affiliation{\como}
\affiliation{\infn}
\affiliation{\NEST}
\author{Giulio Casati}
\email{giulio.casati@uninsubria.it}
\affiliation{\como}
\affiliation{\brazil}
\author{Wen-ge Wang}
\email{wgwang@ustc.edu.cn}
\affiliation{\hefei}

\date{\today}

\begin{abstract}
The correspondence principle is a cornerstone in the entire construction of quantum mechanics. 
This principle has been recently challenged by the observation of an early-time exponential increase of 
the out-of-time-ordered correlator (OTOC) in classically non-chaotic systems
[E.B. Rozenbaum \emph{et al.}, Phys. Rev. Lett. \textbf{125}, 014101 (2020)].
Here we show 
that the correspondence principle is restored after a 
proper treatment of the singular points.
Furthermore our results show that the OTOC maintains its role as a diagnostic of chaotic dynamics.
\end{abstract}
\maketitle
\emph{Introduction.-}
Since the beginning of quantum chaos investigations, 
it has been shown that the exponentially unstable classical motion can persist in quantum mechanics only up to the Ehrenfest time scale $t_E\propto |\ln \hbar|$~\cite{Berman78}, where $\hbar$
is the effective Planck's constant. Indeed, as it was illustrated in~\cite{Casati86}, quantum “chaotic”
motion is dynamically stable. This means that, unlike the exponentially unstable
classical chaotic motion, in the quantum case errors in the initial conditions propagate
only linearly in time. Therefore the quantum diffusion and relaxation process takes
place in the absence of exponential instability,
up to a second, Heisenberg time scale $t_H$ which is 
the minimum time needed to resolve the discretness of 
the operative eigenstates~\cite{qclecture,siberia,qcbook}, namely, those
states which enter the initial conditions and therefore determine the dynamics. 
It should be noticed that, even though the time scale $t_E$ is very short, it diverges as $\hbar$ goes to zero and this ensures the transition to classical motion as required by the correspondence principle.

A popular tool to investigate chaos in quantum systems is the four-point out-of-time-order correlator (OTOC) \cite{Larkin68,Kitaev14,Maldacena14,Maldacena16,Hosur16,Huang16,Swingle16,Galitski17,Yoshii17,Prosen17,Fan17,Garttner17,Li17,Cotler18,Lin18,Jalabert18,Fazio18,Wei18,Dhar18,Richter18,Sondhi18,Nahum18,Khemani18,Rakovszky18,Hirsch19,Borgonovi19,Lakshminarayan19,Carlo19,Hirsch19b,Jalabert19,Wei19,Cory20,Cao20,Xu20,Bunimovich20,Zhou,Ueda18,Jiaozi20}, which can be defined as the expectation of the square
commutator of two operators taken at different times:
\be
{\cal C}(t)=\langle|[\hat{A}(t),\hat{B}(0)]|^{2}\rangle.
\ee
In relation to OTOC, classical and quantum maps and two dimensional billiards have
been studied in great detail \cite{Galitski17,Yoshii17,Cotler18,Zhou,Jalabert18,Ueda18}, since they are more easily amenable to theoretical
and numerical investigations. The importance of these studies is in that, despite their
simplicity, these models exhibit the typical properties of classical and quantum chaos
in more general systems.

The analysis of these systems
 has shown that the short time behaviour of OTOC exhibits
an exponential increase at a rate which is twice the 
Lyapunov exponent of the
corresponding classical system. Quite obviously, for integrable systems, or more
generally for systems with only linear instability, the initial correspondence between
classical and quantum mechanics extends over much longer times.

A recent interesting paper~\cite{Bunimovich20} has introduced a new element which was previously
overlooked and that seems to cast some doubts on the generality of the above
picture. In that paper, classical and quantum polygonal billiards have been
investigated. While these systems are known to have zero Lyapunov exponent, it has
been found that the corresponding quantum billiards display an initial exponential
increase of the quantum mechanical OTOC that has no origin in the classical
counterpart. Moreover the growth rate appears to increase as $\hbar$ is decreased. On the
other hand, since polygons have zero Lyapunov exponent, then the corresponding
classical OTOC does not grow exponentially at any time. The seemingly unavoidable conclusion is
a breakdown of the correspondence principle.

This conclusion is very surprising to us and somehow hard to accept. Indeed, the
correspondence principle is a fundamental stone of the entire construction of quantum
mechanics. As remarked by Max Jammer \cite{Jammer66} ``\textit{In fact, there was rarely in the history
of physics a comprehensive theory which owed so much to one principle as quantum mechanics owed to Bohr's correspondence principle}". The fundamental implications of this
problem require therefore a deep examination. This is the purpose of the present
paper in which we provide convincing evidence that there is no breakdown of the
correspondence principle: the initial growth of the quantum OTOC goes over
smoothly into the classical one and the agreement takes place up to times which
increase as $\hbar$ goes to zero, in accordance with the correspondence principle.
The OTOC then remains a useful diagnosis of chaotic dynamics,  provided  an  appropriate  average  over  initial  states
is done and singularities in the potential are rounded-off below the scale of Planck’s cell.
Our analysis is based on the triangle map, which exhibits the same qualitative properties,
classical and quantum, of triangular or polygonal billiards (in particular, zero Lyapunov exponent and 
exponential growth of the quantum OTOC), while being much simpler for analytical and numerical investigations. 
This is crucial in our case where, in order to discuss the classical
limit, it is desirable to consider sufficiently small values of $\hbar$.


Before discussing our results, we would like to remark that, while the correspondence principle maintains its validity,  the analysis of Ref.~\cite{Bunimovich20} allows to discover an important feature of the
quantum to classical transition. Indeed, polygons have zero Lyapunov exponent but a
round-off of a corner, no matter how small, might lead to chaotic exponentially
unstable motion. On the other hand, in a non-convex polygon, due to the finite size of the quantum
packet, the quantum system will always “see” a rounded vertex and therefore will
move as in a classically chaotic system. This is the reason of the observed initial
exponential growth in non-convex polygons. The importance of this observation is that a similar
phenomenon can take place in generic Hamiltonian systems due to presence of
unstable fixed points which might lead to exponential increase of OTOC even in
integrable systems. This fact may render very delicate the role of OTOC in
discriminating integrable from chaotic systems. Indeed, in recent papers~\cite{Hirsch19b,Cao20} it has
been claimed that exponential growth of OTOC does not necessitate chaos.

\emph{Exponential instability in the round-off triangle map.-}
The triangle map~\cite{Prosen00} is 
defined on the torus with coordinates $(x,p)\in[-1,1)\times[-1,1)$ as follows:
\be
\begin{cases}\label{eq-trmap}
	p_{n+1}=p_{n}-V'(x_n)
& (\text{mod\,2}),\\
	x_{n+1}=x_{n}+p_{n+1} & \text{(mod\,2)},
\end{cases}
\ee
where $V(x)=-\alpha|x|-\beta$.
It is an area preserving, parabolic, piecewise linear map which corresponds to a
discrete bounce map for the billiard in a triangle. The map is marginally 
stable, i.e.,
initially close trajectories separate linearly with time. Even though
the Lyapunov exponent is zero, numerical evidence indicates that this map, for generic, independent irrationals, $\alpha$ and $\beta$ is ergodic and mixing~\cite{Prosen00}.
Hereafter we consider $\beta=0$ for simplicity.
In this latter case the map is only ergodic~\cite{Prosen00}.

\begin{figure}
	\includegraphics[width=1\columnwidth]{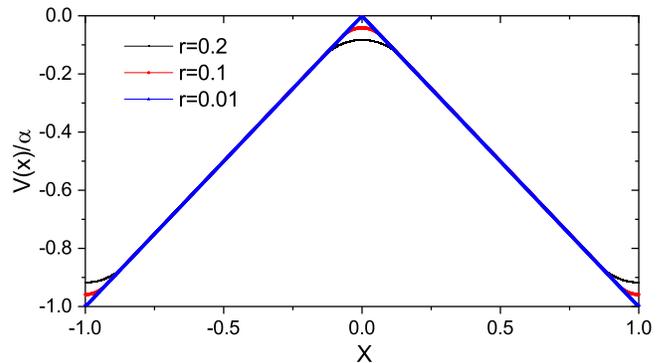}
	\caption{Shape of the potential $V(x)$ for the round-off triangle map, at
different values of $r$.}\label{fig-Vx}
\end{figure}

We also consider the round-off triangle map,
where we substitute the cusps in the potential 
$V(x)$ by small circle arcs of radius $r$
(see Fig.\ref{fig-Vx}):
\be\label{eq-RTM-Vx}
\frac{V(x)}{\alpha}=\begin{cases}
        -\sqrt{2}r+\sqrt{r^{2}-x^{2}}& |x|\le\frac{\sqrt{2}}{2}r,\\
        -1+\sqrt{2}r-\sqrt{r^{2}-(|x|-1)^{2}}&|x|\ge 1-\frac{\sqrt{2}}{2}r,\\
        -|x|&{\rm otherwise}.
\end{cases}
\ee
The original triangle map is recovered for $r=0$.

\begin{figure}
	\includegraphics[width=1\columnwidth]{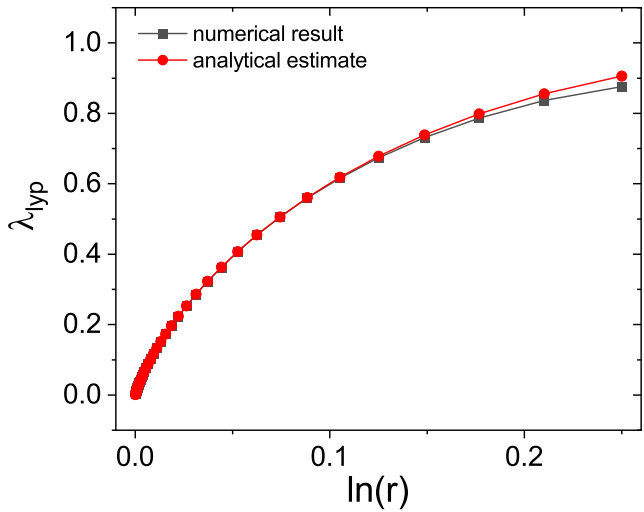}
	\caption{Lyapunov Exponent $\lambda_{\rm lyp}$ as a function of $r$.
The analytical estimate 
of Eq.~(\ref{eq:lyp-new}) is compared with the numerical results. 
Here and in the following figures, $\alpha=[(\sqrt{5}-1)/2-e]/2$.}
\label{le-vs-r}
\end{figure}

The round-off triangle map is exponentially unstable for any $r\ne 0$. 
The Lyapunov exponent can be estimated by considering the tangent map. The length of the tangent vector 
$\left(\begin{array}{c}
        \delta x_{n}\\
        \delta p_{n}
        \end{array}\right)$
increases significantly only when a trajectory reaches the 
neighborhood of $x=0$ or $|x|=1$, that is,
when $|x|<\frac{\sqrt{2}}{2}r$ or $|x|>1-\frac{\sqrt{2}}{2}r$.
We denote these two regions by ${E}_0$ and 
${E}_{|1|}$, 
respectively, and $E=
E_{0}\cup E_{|1|}$.
The width of $E$ is 
$w_E=2\sqrt{2}r$, so that 
the average time between consecutive passages of a trajectory through
$E$ is $\overline{\tau} \simeq  \frac{\sqrt{2}}{2r}$.
For consecutive passages at time steps $t=n$ and $t=n+\tau$, in the case 
of small $r$ we have~\cite{supp}
\be
\left(\begin{array}{c}
        \delta x_{n+\tau}\\
        \delta p_{n+\tau}
\end{array}\right)=\left[\begin{array}{cc}
        \frac{\sqrt{2}\alpha}{r} & \frac{\sqrt{2}\alpha}{r}(\tau-1)\\
        \frac{\sqrt{2}\alpha}{r} & \frac{\sqrt{2}\alpha}{r}(\tau-1)
\end{array}\right]\left(\begin{array}{c}
        \delta x_{n}\\
        \delta p_{n}
\end{array}\right).
\ee
Given the distribution of return times $\tau$ to the region $E$, 
$P(\tau)=q_{r}^{\tau-1}p_{r}$, where $p_{r}=w_E/2=\sqrt{2}r$ and  $q_{r}=1-p_{r}$, we obtain~\cite{supp} 
\begin{align}
\lambda_{\text{lyp}} & =\frac{\sum_{\tau=1}^{\infty}q_{r}^{\tau-1}p_{r}\ln\left(\frac{\sqrt{2}\alpha}{r}\tau\right)}{\overline{\tau}}\nonumber \\
&=2r^{2}\sum_{\tau=1}^{\infty}(1-\sqrt{2}r)^{\tau-1}\ln\left(\frac{\sqrt{2}\alpha}{r}\tau\right).\label{eq:lyp-new}
\end{align}
Note that $\lambda_{\rm lyp}$ decreases with $r$, and $\lambda_{\rm lyp}\to 0$ 
when $r\to 0$.
As shown in Fig.~\ref{le-vs-r}, 
this analytical estimate is in very good agreement 
with the numerically computed Lyapunov exponent.

Similarly, we can also estimate the largest \emph{local} Lyapunov exponent $\lambda^{\text{max}}_{\rm lyp}$
for the region $E$:
\be\label{eq-le-max}
\lambda_{\rm lyp}^{\text{max}}=\ln\left(\frac{\sqrt{2}\alpha}{r}\right).
\ee
In contrast with the 
Lyapunov exponent, $\lambda^{\text{max}}_{\rm lyp}$ increases 
as $r$ decreases.
If we consider the dynamics up to time $t$, the proportion of trajectories 
that satisfy  $x(t')\in E$ for all $t'$ up to some time $t$
is equal to $(\sqrt{2}r)^{t}$,
i.e., it decreases exponentially with time.


\emph{Exponential growth of OTOC.-}
In order to study the quantum evolution we consider the Floquet operator
\be
U=\exp\left(-i\frac{\hat{p}^{2}}{2\hbar}\right)
\exp\left(-i\frac{V(\hat{x})}{\hbar}\right),
\ee
where $\hbar=\frac{2}{\pi D}$, $D$ being the Hilbert space dimension. 
Here we consider the averaged OTOC defined as follows \cite{XP-def-q}:
\be
AL_{q}(t)=\frac{1}{N}\sum_{k=1}^{N}\ln\left(\langle\psi_{k}|[\hat{x}(t),\hat{p}(0)]^{2}|\psi_{k}\rangle\right),
\label{eq:OTOCSchwinger}
\ee
where $|\psi_k\rangle$ is the initial coherent state, which, in the position basis, reads as follows:
\be
\psi_{k}(x)=(\pi\hbar)^{-1/4}\exp\left(-\frac{(x-x_{k})^{2}}{2\hbar}
+\frac{ip_{k}x}{\hbar}\right).
\ee
Here, $(x_k,p_k)$ is the center of the k'th initial state.

\begin{figure}
	\includegraphics[width=0.95\columnwidth]{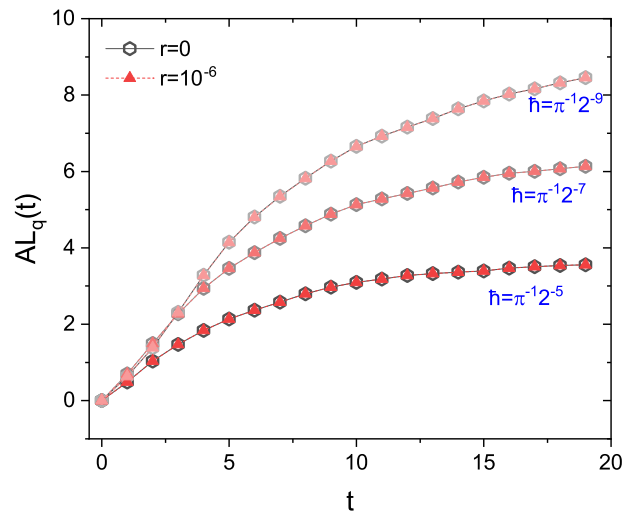}
	\caption{Average OTOC $AL_q(t)$ for different values of $\hbar$, at $r=0$ and $10^{-6}$. 
It can be seen that, for the values of $\hbar$ here considered, 
the growth rate increases with decreasing $\hbar$.
Data for $r=10^{-6}$ are almost indistinguishable from those at $r=0$, as expected since 
in quantum mechanics the sharp, non-analytic features of the triangle-map potential $V(x)$ are
smoothed.}
	\label{AL-h-large}
\end{figure}

In Fig.~\ref{AL-h-large}, we show that the initial growth of the quantum OTOC is exponential
also for the classically non-chaotic case ($r=0$), for which the Lyapunov exponent 
$\lambda_{\rm lyp}=0$. 
Therefore, like in polygons, one can observe that quantum mechanics induces
short-time exponential instability in a classically non-chaotic model.
We plot in Fig.~\ref{AL-h-large} the quantum OTOC for three different values of
$\hbar$. We can see that, in agreement with Fig.~4 of Ref.~\cite{Bunimovich20}, the quantum OTOC
grows exponentially with a rate which increases as $\hbar$ is decreased. This appears to
be the \textit{experimentum crucis} which proves the breakdown of the correspondence
principle. However, as we shall discuss below, a deeper analysis leads 
to a quite different conclusion.

\emph{Quantum-to-classical correspondence.-}
In order to study the quantum-to-classical correspondence, we consider the canonical substitution 
$[\hat{x}(t),\hat{p}(0)]\to i\hbar \{x(t),p(0)\}_{\rm PB}$, where ${\rm PB}$ stands for Poisson brackets. 
We thus obtain the classical counterpart of OTOC as~\cite{XP-def-cl,note:tan}
\be\label{eq-altan}
AL_{c}^{\rm tan}(t)=\frac{1}{N}\sum_{k=1}^{N}\ln\left[\int d\boldsymbol{\gamma}\rho_{\boldsymbol{\gamma}_{0}^{k}}(\boldsymbol{\gamma})\left(\frac{\partial x(t)}{\partial x(0)}\right)^{2}\right],
\ee
where $\gamma=(x,p)$. The initial condition is a Gaussian distribution
\be\label{eq:InitialC}
\rho_{\boldsymbol{\gamma_{0}^{k}}}(\boldsymbol{\gamma})=(2\pi\sigma^{2})^{-1}
\exp\left(-\frac{(x-x_{k})^{2}+(p-p_{k})^{2}}{2\sigma^{2}}\right),
\ee
where, in order to compare with the quantum wavepacket, 
we take $\sigma=\sqrt{\frac{\hbar_c}{2}}$ and ${\hbar_c}=\hbar$.

There are obviously no problems for the correspondence principle when $r>0$. 
The round-off triangle map is chaotic and therefore one expects that classical and
quantum OTOC agree up to the Ehrenfest time. This is nicely confirmed by our
numerical computations (see~\cite{supp}) where one can see that the 
growth rate of $AL^{\rm tan}_c(t)$ approaches
the Lyapounov exponent as $\hbar$ goes to zero while $AL_q(t)$ approaches the
corresponding $AL^{\rm tan}_c(t)$ up to the Ehenrefest time. 

On the other hand, the case $r=0$ requires careful inspection. First of all, we observe that there are
singular points (the cusps in the potential) for which $\partial x(t)/\partial x(0)$ diverges. 
This leads to a divergence of the growth rate for $AL_{c}^{\rm tan}$, as explained in what follows.

Besides $AL_{c}^{\rm tan}$, we consider two other ways of averaging over initial conditions:
	\be
	LA_{c}^{\rm tan}(t)=\ln\left[\frac{1}{N}\sum_{k=1}^{N}\int d\boldsymbol{\gamma}\rho_{\boldsymbol{\gamma}_{0}^{k}}(\boldsymbol{\gamma})\left(\frac{\partial x(t)}{\partial x(0)}\right)^{2}\right],
	\ee
	and
	\be
	LL_{c}^{\rm tan}(t)=\frac{1}{N}\sum_{k=1}^{N}\int d\boldsymbol{\gamma}\rho_{\boldsymbol{\gamma}_{0}^{k}}(\boldsymbol{\gamma})\ln\left(\frac{\partial x(t)}{\partial x(0)}\right)^{2}.
	\ee
We can see that $LL_{c}^{\rm tan}$ is an average of the quantity considered in computing the Lyapunov exponent. For a number $N$ of initial 
conditions large enough, we can expect that 
\be
	LL_{c}^{\rm tan}(t)\propto 2\lambda_{\text{\rm lyp}}t.
	\ee
	As for $LA_{c}^{\rm tan}(t)$, it is close to $LL_{c}^{\rm tan}(t)$ only if the fluctuations from trajectory to trajectory of the local Lyapunov exponent are quite small.  On the other hand, for small $r$ such fluctuations are large and $LA_{c}^{\rm tan}(t)$ is dominated by the trajectories with largest local Lyapunov exponent $\lambda^{\text{max}}_{\rm lyp}$.
	Given  $\lambda_{\rm lyp}^{\text{max}}$ from Eq.~(\ref{eq-le-max}), and the fraction $(\sqrt{2}r)^{t}$ of the trajectories with largest local Lyapunov 
	exponent up to time $t$, we conclude that, for $r\to 0$,
    \be
    LA_{c}^{\rm tan}(t)\propto 2\lambda_{\rm lyp}^{*}t,
    \ee
    where
    \be\label{eq-lambdas}
    \lambda_{\rm lyp}^{*}\approx \ln\left(\frac{\sqrt{2}\alpha}{r}\right)+\frac{1}{2}\ln(\sqrt{2}r).
    \ee
    Therefore, the growth rate of $LA_{c}^{\rm tan}$ diverges when $r\to 0$, in spite of the fact that in that limit the system is 
    classically integrable. 
    		
	Then we come to the discussion of the quantity $AL_{c}^{\rm tan}(t)$.  
	For sufficiently small $\hbar_c$, for each single initial ensemble at small times all the trajectories remain very close, at distances much smaller than $r$. Then the behavior of $AL_{c}^{\rm tan}(t)$ is quite similar to that of $LL_{c}^{\rm tan}(t)$. On the other hand, for longer times, when the size of the ensemble becomes much larger than $r$, $AL_{c}^{\rm tan}(t)$ is close to $LA_{c}^{t\rm an}(t)$. 
	As a conclusion, for $\sigma=\sqrt{\frac{\hbar}{2}}\ll r$ we obtain
	\be
	AL_{c}^{\rm tan}(t)\propto\begin{cases}
		LL_{c}^{\rm tan}(t)\propto 2\lambda_{\text{lyp}}t & t\ll t^{*},\\
		LA_{c}^{\rm tan}(t)\propto 2\lambda_{\text{lyp}}^{*}t & t\gg t^{*},
	\end{cases}
	\ee
	where $t^{*}$ indicates the time scale when the size  $\Delta X(t)\sim \sqrt{\hbar_c} \exp{(\lambda_{\text{lyp}} t)}$
	of the wave packet becomes comparable with $r$.
Therefore, we can estimate the value of $t^*$  as
    \be\label{eq:ts-origin}
    t^{*}\sim \frac{1}{\lambda}\ln\frac{r}{\sqrt{\hbar_{c}}}.
    \ee
   For a fixed $r$,  when $\hbar_c$ is large $t^*$ is very small, and $AL^{\text{\rm tan}}_c(t)$ 
   increases with growth rate $2\lambda^*_{\text{lyp}}$. 
   On the other hand, for $\hbar_c\rightarrow 0$ we have  $t^*\rightarrow \infty$, 
   and the initial growth rate is given by $2\lambda_{\text{lyp}}$ (see~\cite{supp} for numerical confirmation of the above picture).

\begin{figure}
	\includegraphics[width=1\columnwidth]{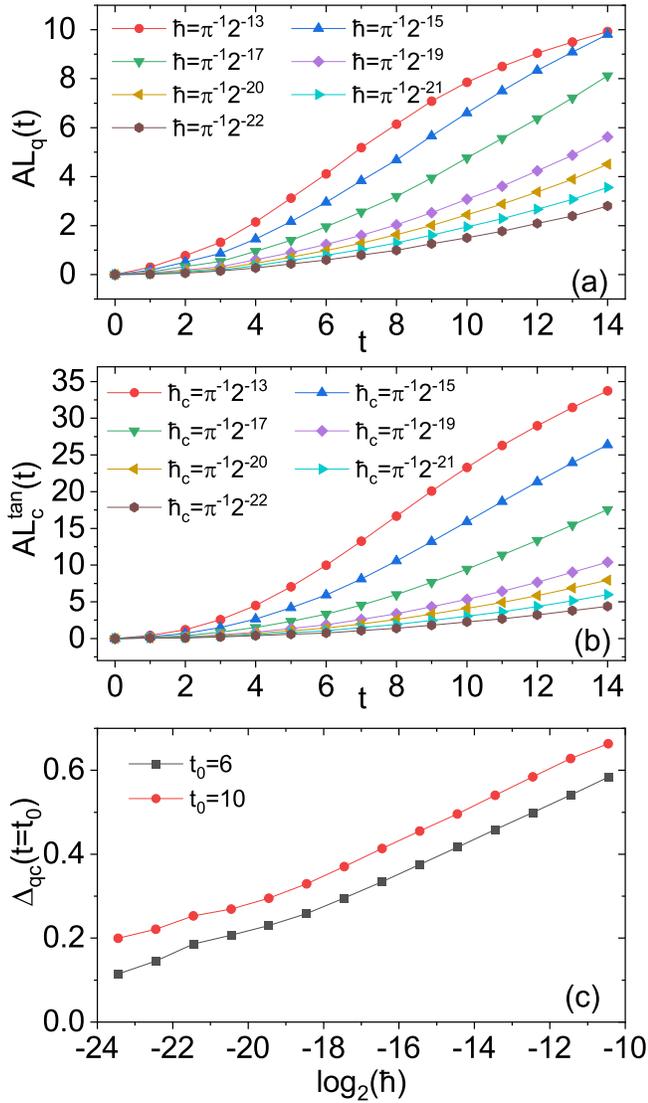}
        \caption{$AL(t)$ in (a) quantum case for $r=0$ and (b) its classical counterpart for different $\hbar$  from $\hbar=\pi^{-1}2^{-13}$ to  $\hbar=\pi^{-1}2^{-22}$.
                The classical counterpart $AL^{\text{tan}}_c(t)$ is obtained by considering a finite $r$, which is related to the dimension $D$ of the system as $r=\frac{1}{\sqrt{D}}$.
            (c) Difference between $AL_q(t)$ and  $AL^{\text{tan}}_c(t)$ at a fixed time $t_0$, denoted by $\Delta_{qc} (t=t_0)$, where $t_0=6$ (black blocks) and $t_0=10$ (red circles). }\label{AL-r0-qc}
\end{figure}

To examine the validity of the correspondence principle for $r=0$, we first compute the quantum OTOC $AL_q(t)$ at different values of $\hbar$.
 Numerical results are shown in Fig.~\ref{AL-r0-qc}(a). It is clear that the growth rate of $AL_q$, which we have seen 
 to increase with decreasing $\hbar$ (down to $\hbar=\pi^{-1}2^{-9}$), vanishes instead when $\hbar\to 0$, in accordance with the 
 correspondence principle.
For a detailed classical-quantum comparison, given that quantum mechanics smoothens the sharp features of the classical potential 
below the Planck's scale, we juxtapose
 the quantum results for OTOC at $r=0$ with the classical ones at $r=1/\sqrt{D}$.
  As shown in Fig.~\ref{AL-r0-qc}(b), also the growth rate of $AL^{\rm tan}_c$ vanishes when $\hbar_c\to 0$. 
  In order to get a clear picture of the difference between the quantum and classical results, we consider the relative difference
 of $AL_q$ and $AL_c^{\rm tan}$,
 \be
 \Delta_{qc}(t) =|AL_q (t)-AL^{\rm tan}_c(t)|/[AL_q (t)+AL^{\rm tan}_c(t)].
 \ee
 The results for $t=t_0=6,10$ are shown in Fig.~\ref{AL-r0-qc}(c).
 It is clear that in both cases $\Delta_{qc}\to 0$  with decreasing $\hbar$.
These results  show that there is no breakdown of the correspondence principle.


It is intriguing that the OTOC growth rate exhibits a nonmonotonous dependence on $\hbar$. 
While we do not have a rigorous explanation for this numerical result, a possible clue is the following.
Due to the finite size of the wave packet, the quantum system “sees” a rounded potential, with effective radius $r=f(\hbar)$, where $f$
is a monotonous growing function of $\hbar$. We then have, as we have discussed for the classical case, a growth rate $2\lambda_{\rm lyp}$ 
up to a time $t^\star$ and then a growth rate $2\lambda_{\rm lyp}^\star$. Numerical data as well as Eq.~(\ref{eq:ts-origin}) suggest that $t^\star$ increases when $\hbar$ decreases,
in such a way that the initial growth rate is determined by the fluctuations in the local Lyapunov exponent for large $\hbar$, and by the 
Lyapunov exponent for small $\hbar$. 
In particular, the OTOC growth rate in Fig.~\ref{AL-h-large} is not given by the Lyapunov exponent.

\emph{Conclusions.-}
In recent years, the OTOC has emerged as an important tool to characterize chaos in many-body quantum systems. 
His validity, first corroborated by models which exhibit an exponential increase of OTOC with rate equal to twice the Lyapunov 
exponent of the underlying classical dynamics, has been more recently questioned. Indeed, unstable fixed points might lead to
an exponential increase of OTOC even in integrable systems~\cite{Hirsch19b,Cao20}. 
Even more importantly, the exponential increase can be observed at early times, 
questioning the validity of the correspondence principle~\cite{Bunimovich20}. 
Our results show that the correspondence principle is restored and the OTOC remains a useful diagnosis of chaotic dynamics, 
provided an appropriate average over initial states is done and singularities in the potential are rounded-off below the scale 
of Planck's cell.

{\it Acknowledgments}:
J.W. and W-G.W. acknowledge 
the Natural Science Foundation of China under Grant
Nos.~11535011, and 11775210.
G.B. acknowledges the financial support of the INFN through the project “QUANTUM”.


\begin{appendix}

\section{Quantum Chaos and the Correspondence Principle: Supplemental Material}

\section{Analytical estimate of the Lyapunov exponent}

	
We consider the tangent map for the round-off triangle map:
        \be\label{eq-tan-RTM}
\left(\begin{array}{c}
        \delta x_{n+1}\\
        \delta p_{n+1}
\end{array}\right)=\left[\begin{array}{cc}
        1-V''(x_n) & 1\\
        -V''(x_n) & 1
\end{array}\right]\left(\begin{array}{c}
        \delta x_{n}\\
        \delta p_{n}
\end{array}\right),
        \ee
where
	\be\label{eq-VX2-RTM}
	V''(x)=\begin{cases}
		-\frac{\alpha}{\sqrt{r^{2}-x^{2}}}-\frac{\alpha x^{2}}{(\sqrt{r^{2}-x^{2}})^{3}} & |x|\le\frac{\sqrt{2}}{2}r\\
		\frac{\alpha}{\sqrt{r^{2}-(x-1)^{2}}}+\frac{\alpha(x-1)^{2}}{(\sqrt{r^{2}-(x-1)^{2}})^{3}} & x\ge1-\frac{\sqrt{2}}{2}r\\
		\frac{\alpha}{\sqrt{r^{2}-(x+1)^{2}}}+\frac{\alpha(x-1)^{2}}{(\sqrt{r^{2}-(x+1)^{2}})^{3}} & x\le-1+\frac{\sqrt{2}}{2}r\\
		0 & {\rm otherwise}.
	\end{cases}
	\ee
	
				\begin{figure}
		\includegraphics[width=1\columnwidth]{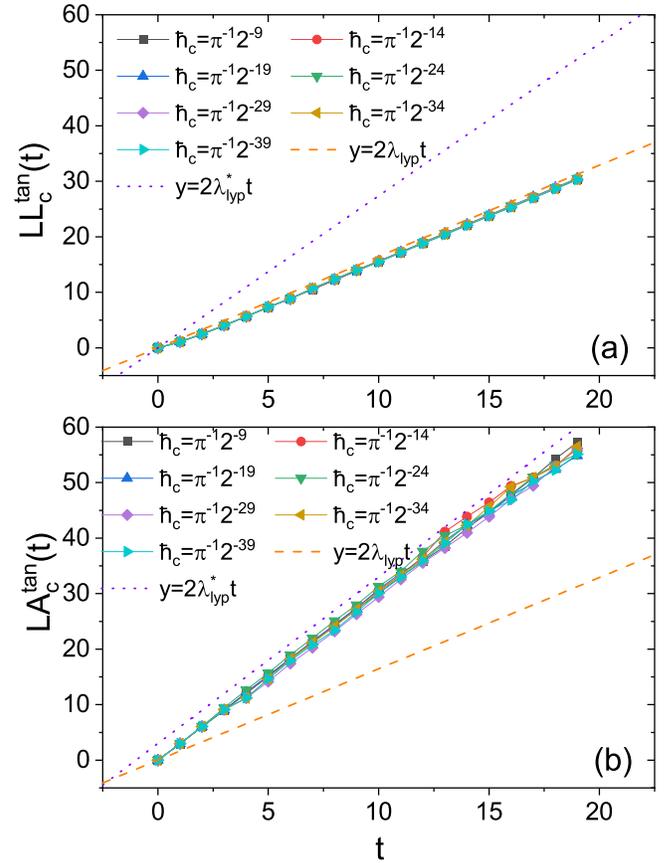}
		\caption{(a)$LL^{\rm tan}_c(t)$ and (b)$LA^{\rm tan}_c(t)$ for $r=0.2$ and different $\hbar_c$  from $\hbar_c=\pi^{-1}2^{-9}$ to  $\hbar_c=\pi^{-1}2^{-39}$; the curves $y=2\lambda_{\text{lyp}}t$ and  $y=2\lambda^*_{\text{lyp}}t$ are shown for comparison. Here $\lambda_{\text{lyp}}$ is given by Eq.(\ref{eq:lyp-new}), while $\lambda^*_{\text{lyp}}$ is given by Eq.~(\ref{eq-lambdas}) of the main text.}
		\label{LAt-LLT}
	\end{figure}

The average value of $V''(x)$ for  $x\in E_0$ is 
	\be
	\langle V''(x)\rangle_{E_{0}}=\frac{1}{\sqrt{2}}\int_{-\frac{\sqrt{2}}{2}}^{+\frac{\sqrt{2}}{2}} V''(x) dx=-\frac{\sqrt{2}\alpha}{r}.
	\ee
	Similarly, one has
	\be
\langle V''(x)\rangle_{E_{1}}=
\frac{\sqrt{2}\alpha}{r}.
	\ee
For small $r$, the length of the tangent vector 
increases significantly only when $x\in E$. 
	We first consider the time step $t=n$, 
when the trajectory reaches $E$ for the first time.
For small $r$, one has the following approximation
	of the tangent map:
	\be
\left(\begin{array}{c}
	\delta x_{n}\\
	\delta p_{n}
\end{array}\right)=\left[\begin{array}{cc}
	\frac{\sqrt{2}\alpha}{r} & 1\\
	\frac{\sqrt{2}\alpha}{r} & 1
\end{array}\right]\left(\begin{array}{c}
	\delta x_{n-1}\\
	\delta p_{n-1}
\end{array}\right).
	\ee
	We then consider the time step $t=n+\tau$, at which the trajectory 
reaches $E$ for a second time. We have
\be
\left(\begin{array}{c}
	\delta x_{n+\tau}\\
	\delta p_{n+\tau}
\end{array}\right)=\left[\begin{array}{cc}
	\frac{\sqrt{2}\alpha}{r} & 1\\
	\frac{\sqrt{2}\alpha}{r} & 1
\end{array}\right]\left[\begin{array}{cc}
	1 & 1\\
	0 & 1
\end{array}\right]^{\tau-1}\left(\begin{array}{c}
	\delta x_{n}\\
	\delta p_{n}
\end{array}\right),
\ee
which leads to
\be\label{eq:mapMatrix}
\left(\begin{array}{c}
	\delta x_{n+\tau}\\
	\delta p_{n+\tau}
\end{array}\right)=\left[\begin{array}{cc}
	\frac{\sqrt{2}\alpha}{r} & \frac{\sqrt{2}\alpha}{r}(\tau-1)\\
	\frac{\sqrt{2}\alpha}{r} & \frac{\sqrt{2}\alpha}{r}(\tau-1)
\end{array}\right]\left(\begin{array}{c}
	\delta x_{n}\\
	\delta p_{n}
\end{array}\right).
\ee
Replacing $\tau$ by the average time $\overline{\tau}$ between consecutive passages of a trajectory through $E$, we can
estimate the Lyapunov exponent of the system as 
\be
\lambda_{{\rm lyp}}=\frac{\ln(\frac{\sqrt{2}\alpha}{r}\overline{\tau})}{\overline{\tau}}=\sqrt{2}r\ln\left(\frac{\alpha}{r^{2}}\right).
\label{eq:lyap}
\ee

In order to get a more accurate estimate, 
we study the distribution of the return times $\tau$, $P(\tau)=q_{r}^{\tau-1}p_{r}$, where $p_{r}=\sqrt{2}r$ and $q_{r}=1-p_{r}$.
Then, it should be noticed here that the mapping matrix in Eq~.(\ref{eq:mapMatrix}) belong 
to a special class of matrices which can be written in the following form:
\be
\hat{M}(a,b)=\left[\begin{array}{cc}
	a & b\\
	a & b
\end{array}\right].
\ee
Considering a sufficiently long time, when the trajectory reaches $E$ for $m_0$ times,
the Lyapunov exponents can be estimated as follows:
\be
\lambda_{\text{lyp}}=\frac{\ln E_{\rm max}(\hat{M}_{m_{0}})}{m_{0}\tau},
\ee 
where
\be
\hat{M}_{m_{0}}=\prod_{k=1}^{m_{0}}\hat{M}\left(\frac{\sqrt{2}\alpha}{r},\frac{\sqrt{2}\alpha}{r}(\tau_{k}-1)\right)
\ee
and $E_{\rm max}(\hat{M})$ is the maximum eigenvalue of $\hat{M}$.
We note that matrices $\hat{M}(a,b)$ have the property 
\be
E_{\rm max}\left(\prod_{k=1}^{n}\hat{M}(a_{k},b_{k})\right)=\prod_{k=1}^{n}(a_{k}+b_{k}).
\ee
Considering this expression, together with $P(\tau)$, we obtain the value of 
$\lambda_{{\rm lyp}}$ reported in the main text.



	\begin{figure}
	\includegraphics[width=1\columnwidth]{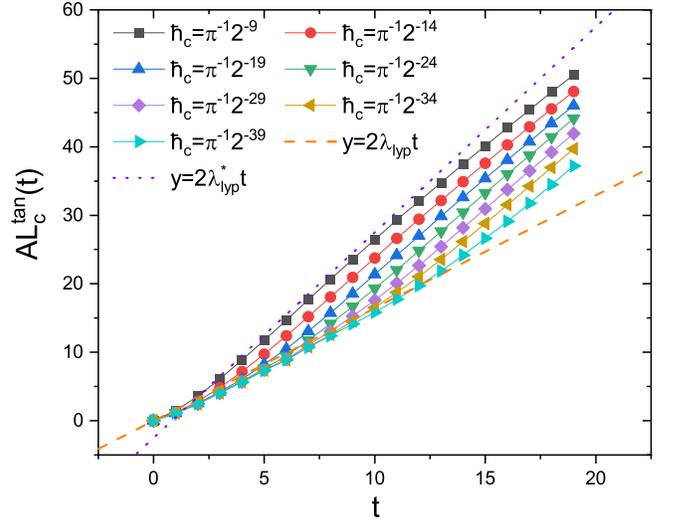}
	\caption{$AL^{tan}_c(t)$ for the same parameter values as in Fig.~\ref{LAt-LLT}. }
	\label{AL-r02}
\end{figure}

	\section{Comparison of different ways of averaging}
	In this section, we show numerical results confirming the analytical predictions discussed in the main text for the 
 different ways of averaging the classical OTOC over initial conditions:
	\be
	LL_{c}^{\rm tan}(t)=\frac{1}{N}\sum_{k=1}^{N}\int d\boldsymbol{\gamma}\rho_{\boldsymbol{\gamma}_{0}^{k}}(\boldsymbol{\gamma})\ln\left(\frac{\partial x(t)}{\partial x(0)}\right)^{2},
			\label{eq:LLc}
	\ee
		\be
	LA_{c}^{\rm tan}(t)=\ln\left[\frac{1}{N}\sum_{k=1}^{N}\int d\boldsymbol{\gamma}\rho_{\boldsymbol{\gamma}_{0}^{k}}(\boldsymbol{\gamma})\left(\frac{\partial x(t)}{\partial x(0)}\right)^{2}\right],
		\label{eq:LAc}
	\ee
			\be
	AL_{c}^{\rm tan}(t)=\frac{1}{N}\sum_{k=1}^{N}\ln\left[\int d\boldsymbol{\gamma}\rho_{\boldsymbol{\gamma}_{0}^{k}}(\boldsymbol{\gamma})\left(\frac{\partial x(t)}{\partial x(0)}\right)^{2}\right],
	\label{eq:ALc}
	\ee

\begin{figure}
	\includegraphics[width=1\columnwidth]{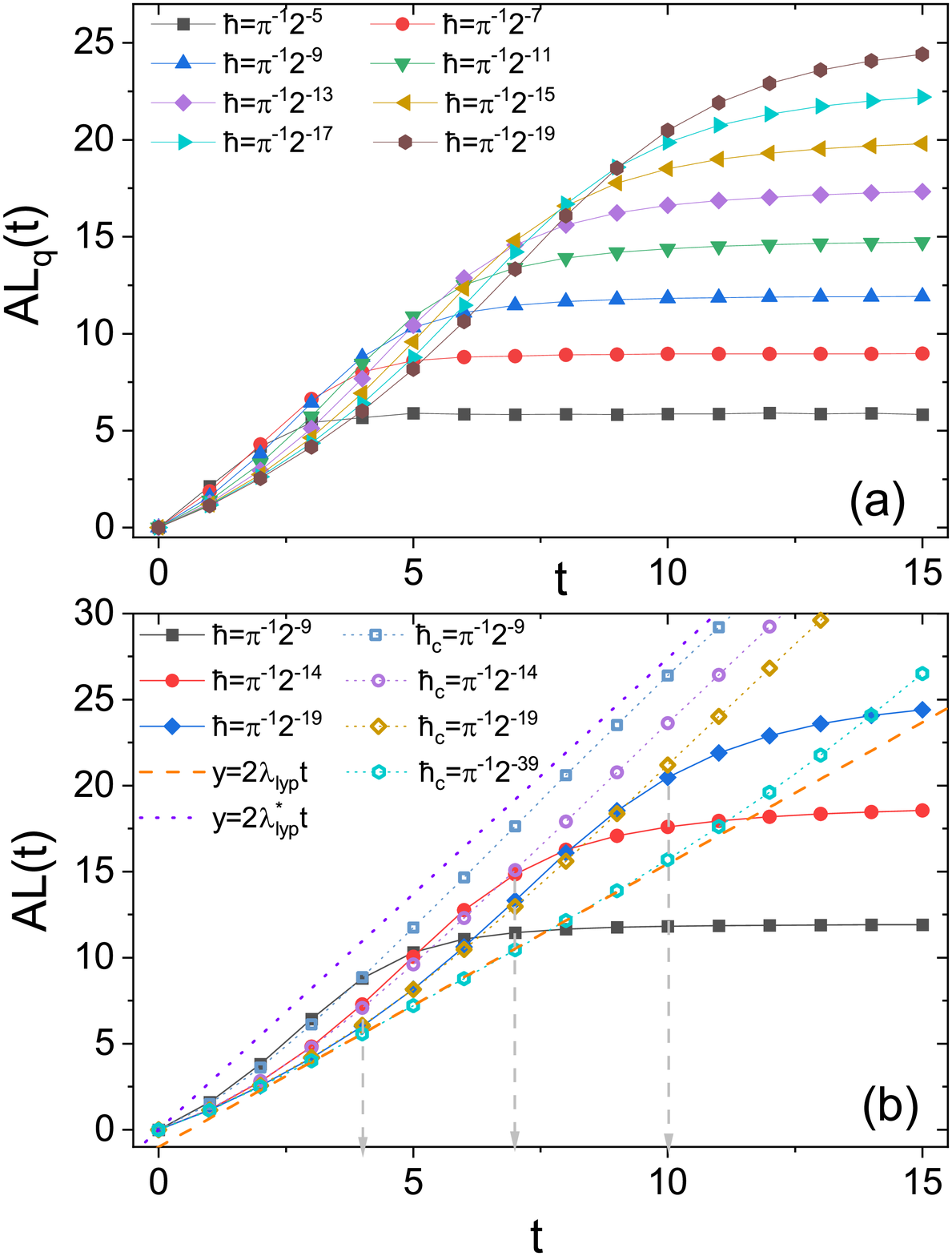}
	\caption{(a) $AL_q(t)$ in the quantum case  for different $\hbar$, and  $r=0.2$. (b) Comparison of $AL(t)$ in quantum (solid symbols) and 
		classical (open symbols) case for $r=0.2$.  
		The curves $y=2\lambda_{\text{lyp}}t$ and $y=2\lambda_{\text{lyp}}^\star$ are also shown for comparison. 
		The Ehrenfest times at different $\hbar$ are marked by vertical gray dashed lines.
	}
	\label{AL-h-all}
\end{figure}

	Numerical results for $LL_{c}^{\rm tan}$ and $LA_{c}^{\rm tan}$ are shown in Fig.~\ref{LAt-LLT}, for $r=0.2$ and different values of 
	$\hbar_c$, from $\pi^{-1}2^{-9}$ to  $\pi^{-1}2^{-39}$.  The agreement between the numerical growth rates and those derived analytically in the Eq.(\ref{eq:lyp-new}) and Eq.(\ref{eq-le-max}), 
	$\lambda_{\rm lyp}$ and $\lambda_{\rm lyp}^\star$ respectively, is excellent. Moreover, data over a broad range of values of $\hbar_c$ collapse. 
	The independence of the growth rate on the of the initial distribution, of variance $\sigma=\sqrt{\frac{\hbar_c}{2}}$, can be understood as follows. 
	The centers of the $N$ initial conditions evolve in time and rapidly distribute uniformly in the phase space. If $N$ is large enough, we 
	can substitute the integrals in (\ref{eq:LLc}) and (\ref{eq:LAc}) with the average over the whole phase space:
		\begin{gather}
	LL_{c}^{tan}(t)\simeq\frac{1}{\Pi}\int dxdp\ln\left(\frac{\partial x(t)}{\partial x(0)}\right)^{2}, \label{eq-LL1} \\
	LA_{c}^{tan}(t)\simeq\ln\left[\frac{1}{\Pi}\int dxdp\left(\frac{\partial x(t)}{\partial x(0)}\right)^{2}\right], \label{eq-LA1}
	\end{gather}
	where $\Pi$ is the volume of the whole phase space.

	For $AL_c^{\rm tan}$, as discussed in the main text we expect the  growth rate $2\lambda_{\text{lyp}}$ for $t<t^\star\sim (1/\lambda)\ln (r/\sqrt{\hbar_c})$,
	and the growth rate  $2\lambda_{\text{lyp}}^\star$ for $t>t^\star$. Such expectations are clearly confirmed by our numerical data, shown for $r=0.2$ in Fig.~\ref{AL-r02}.

\section{Ehrenfest time scale}

In this section we show an example of quantum-classical corresponcence for the chaotic, round-off triangle map ($r>0$).
We can see in Fig.~\ref{AL-h-all} that both the classical and the quantum OTOC  grow initially with rate $2\lambda_{\text{lyp}}$ and then, 
after a time $t^\star$ discussed in the main text, with rate $2\lambda_{\text{lyp}}^\star$. 
The quantum OTOC then obviously saturates due to the 
finite size of the Hilbert space.   Classical and quantum OTOC agree up to the Ehrenfest time 
$t_E\propto | \ln \hbar |$, marked in Fig.~\ref{AL-h-all} by gray vertical lines 
for different $\hbar$.
To summarize, the growth rate of $AL_c^{\rm tan}(t)$ approaches the Lyapunov exponent as $\hbar_c$
goes to zero while $AL_q(t)$ approaches the corresponding $AL_c^{\rm tan}(t)$ up to the Ehenrefest time.

\end{appendix}


\begin{thebibliography}{99}
\bibitem{XP-def-q} The operators $\hat{x}$ and $\hat{p}$ used to compute the OTOC are defined as in~\cite{Jalabert18}. Such construction, in terms of Schwinger shifts operators, avoids problems related to the definition of  $\hat{x}$ and $\hat{p}$ on the quantized torus, which is equivalently to consider the operator $\sin \pi \hat{x}$ and   $\sin \pi \hat{p}$ instead. So what is actually studied here is the following quantity: $AL_{q}(t)=\frac{1}{N}\sum_{k=1}^{N}\ln\left(\langle\psi_{k}|[\sin(\pi\hat{x}(t)),\sin(\pi\hat{p}(0)])^{2}|\psi_{k}\rangle\right).$
\bibitem{note:tan} The superscript ``tan'' refers to the fact that in computing the OTOC we use the tangent map rather than evolving two nearby 
trajectories and computing their distance $\delta x(t)$.
\bibitem{XP-def-cl} Given the definition of $\hat{x}$ and $\hat{p}$ explained in \cite{XP-def-q}, 
we actually compute 
$AL_{c}^{\rm tan}(t)=\frac{1}{N}\sum_{k=1}^{N}\ln\left[\pi^{2}\int d\boldsymbol{\gamma}\rho_{\boldsymbol{\gamma}_{0}^{k}}(\boldsymbol{\gamma})\left(\cos(\pi x(t))\cos(\pi p(0))\frac{\partial x(t)}{\partial x(0)}\right)^{2}\right].$
\end{thebibliography}
\end{document}